\documentclass[sigplan,9pt,screen]{acmart}
\pdfoutput=1
\renewcommand\footnotetextcopyrightpermission[1]{}

\usepackage{booktabs}   
\usepackage{subcaption} 
\usepackage{amsmath}
\usepackage{amssymb}
\usepackage{amsthm}
\usepackage{textcomp}
\usepackage{mathtools}
\usepackage{amscd}
\usepackage{amsfonts}
\usepackage{varwidth}
\usepackage{graphicx}
\usepackage{listings}
\usepackage{float}
\usepackage{multirow}
\usepackage[noend]{algorithmic}
\usepackage{mathrsfs}
\usepackage{xspace}
\usepackage{enumitem}

\makeatletter 
\def\arcr{\@arraycr}
\makeatother

\usepackage{flushend}

\usepackage{fancyhdr}
\usepackage{hyphenat}

\usepackage{libertine}
\usepackage[T1]{fontenc}

\setlist[itemize]{topsep=0pt,itemsep=-1ex,partopsep=1ex,parsep=1ex,itemindent=0pt,leftmargin=8pt}

\definecolor{deepblue}{rgb}{0,0,0.5}
\definecolor{deepred}{rgb}{0.6,0,0}
\definecolor{deepgreen}{rgb}{0,0.5,0}
\definecolor{halfgray}{gray}{0.55}
\definecolor{ipythonframe}{RGB}{207, 207, 207}

\definecolor{ckeyword}{HTML}{7F0055}
\definecolor{ccomment}{HTML}{3F7F5F}
\definecolor{cnumber}{HTML}{2A0099}
\definecolor{pblue}{rgb}{0.13,0.13,1}
\definecolor{pgreen}{rgb}{0,0.5,0}
\definecolor{pred}{rgb}{0.9,0,0}
\definecolor{pgrey}{rgb}{0.46,0.45,0.48}

\lstdefinelanguage{Scilla}{ 
  keywords={typeof, for, enum, struct, modifier, function, public,
  returns, external, contract, new, true, then, false, private, catch,
  function, return, null, throw, catch, switch, var, if, in,
  transition, let, while, do, else, case, break, require,
  continuation, send}, ndkeywords={bool, address, mapping, uint,
  bytes32, string, uint256}, 
  identifierstyle=\color{black},
  sensitive=false,
  comment=[l]{//},
  morecomment=[s]{(*}{*)},
  commentstyle=\color{ccomment}\ttfamily,
  string=[b]",
  showstringspaces=false,
  morestring=[b]',
  showspaces=false,
  showtabs=false,
  breaklines=true,
  morekeywords={function, contract, returns, return},
  breakatwhitespace=true,
  lineskip=-0.6pt,
  basewidth={0.54em, 0.4em},%
  basicstyle=\footnotesize\ttfamily,
  keywordstyle={\color{ckeyword}\ttfamily\bfseries},
  ndkeywordstyle={\color{pblue}\ttfamily\bfseries},
  commentstyle={\color{ccomment}\itshape},
  stringstyle={\color{pgreen}\ttfamily},
  numberstyle={\scriptsize\color{cnumber}\sffamily},
  moredelim=[il][\textcolor{pgrey}]{$$},
  moredelim=[is][\textcolor{pgrey}]{\%\%}{\%\%},
  literate={<-}{{$\leftarrow$}}1
           {=>}{{$\Rightarrow$}}1
           {->}{{$\rightarrow$}}1
           {<=}{{$\leq$}}1
}

\newcommand{\scode}[1]{\lstinline[language=Scilla,basicstyle=\small\ttfamily]{#1}}
\newcommand{\code}[1]{\scode{#1}}
\newcommand{\ccode}[1]{\lstinline[language=Coq,basicstyle=\small\ttfamily]{#1}}

\definecolor{shadecolor}{gray}{1.00}
\definecolor{ddarkgray}{gray}{0.75}
\definecolor{darkgray}{gray}{0.30}
\definecolor{light-gray}{gray}{0.87}

\newcommand{\etc}{\emph{etc}}
\newcommand{\ie}{\emph{i.e.}\xspace}

\newcommand{\eg}{\emph{e.g.}\xspace}

\newcommand{\etal}{\emph{et~al.}\xspace}

\newcommand{\wrt}{\emph{wrt.}\xspace}

\newcommand{\scilla}{{\sc Scilla}\xspace}

\definecolor{shadecolor}{gray}{1.00}
\definecolor{darkgray}{gray}{0.30}
\definecolor{violet}{rgb}{0.56, 0.0, 1.0}
\definecolor{forestgreen}{rgb}{0.13, 0.55, 0.13}

\lstdefinelanguage{Coq} {
mathescape=true,						
texcl=false,
morekeywords=[1]{
  Add,
  All,
  Arguments,
  Axiom,
  Bind,
  Canonical,
  Check,
  Close,
  CoFixpoint,
  CoInductive,
  Coercion,
  Contextual,
  Corollary,
  Defined,
  Definition,
  Delimit,
  End,
  Example,
  Export,
  Fact,
  Fixpoint,
  Goal,
  Graph,
  Hint,
  Hypotheses,
  Hypothesis,
  Implicit,
  Implicits,
  Import,
  Inductive,
  Lemma,
  Let,
  Local,
  Locate,
  Ltac,
  Maximal
  Module,
  Morphism,
  Next,
  Notation,
  Obligation,
  Open,
  Parameter,
  Parameters,
  Prenex,
  Print,
  Printing,
  Program,
  Projections,
  Proof,
  Proposition,
  Qed,
  Record,
  Relation,
  Remark,
  Require,
  Reserved,
  Resolve,
  Rewrite,
  Save,
  Scope,
  Search,
  Section,
  Show,
  Strict,
  Structure,
  Tactic,
  Theorem,
  Unset,
  Variable,
  Variables,
  View,
  inside,
  outside
},
morekeywords=[2]{
  as,
  cofix,
  else,
  end,
  exists,
  exists2,
  fix,
  for,
  forall,
  fun,
  if,
  in,
  is,
  let,
  match,
  nosimpl,
  of,
  return,
  struct,
  then,
  vfun,
  with
},
morekeywords=[3]{Type, Prop, Set, True, False},
morekeywords=[4]{
  after,
  apply,
  assert,
  auto,
  bool_congr,
  case,
  change,
  clear,
  compute,
  congr,
  cut,
  cutrewrite,
  destruct,
  elim,
  field,
  fold,
  generalize,
  have,
  heval, 
  hnf,
  induction,
  injection,
  intro,
  intros,
  intuition,
  inversion,
  left,
  loss,
  move,
  nat_congr,
  nat_norm,
  pattern,
  pose,
  refine,
  rename,
  replace,
  revert,
  rewrite,
  right,
  ring,
  set,
  simpl,
  split,
  suff,
  suffices,
  symmetry,
  transitivity,
  trivial,
  unfold,
  unlock,
  using,
  without,
  wlog,
  autorewrite
},        
morekeywords=[5]{
  assumption,
  by,
  contradiction,
  done,
  exact,
  lia,
  gappa,
  omega,
  reflexivity,
  romega,
  solve,
  tauto,
  discriminate,
  unsat
},
morecomment=[s]{(*}{*)},
morekeywords=[6]{do, first, try, idtac, repeat},
showstringspaces=false,
morestring=[b]",
tabsize=3,							
extendedchars=true,  		 		
sensitive=true, 
breaklines=false,
basicstyle=\footnotesize\ttfamily,
captionpos=b,							
columns=[l]fullflexible,
identifierstyle={\color{black}},
keywordstyle=[1]{\color{violet}},
keywordstyle=[2]{\color{ckeyword}},
keywordstyle=[3]{\color{ckeyword}},
keywordstyle=[4]{\color{blue}},
keywordstyle=[5]{\color{red}},
keywordstyle=[6]{\color{violet}},
stringstyle=,
commentstyle=\it\ttfamily\color{ccomment},
numberstyle={\scriptsize\color{cnumber}\sffamily},
literate={\\/}{{$\vee$}}1
         {/\\}{{$\wedge$~}}1
         {:->}{{$\mapsto~$\!}}1
         {<--}{{$\asgn~$}}1
         {\\in}{{$\in~$}}1
         {\\In}{{$\in~$}}1
         {++}{{$+\!+\!~$}}1
         {->}{{$\to~$}}1
         {negb}{{$\neg~$}}1
         {forall}{{$\forall~$}}1
         {exists}{{$\exists~$}}1
         {=>}{{$\Rightarrow~$}}1
         {\\+}{{$\!\join\!~$}}1
}

\lstdefinestyle{Coq}{language=Coq}

\begin{document}

\bibliographystyle{plainnat}

\hypersetup{colorlinks,
  linkcolor=ACMDarkBlue,
  citecolor=ACMPurple,
  urlcolor=ACMDarkBlue,
  filecolor=ACMDarkBlue}


\title{\scilla: a Smart Contract Intermediate-Level LAnguage}
\subtitle{Automata for Smart Contract Implementation and Verification}

\author{Ilya Sergey}
\affiliation{
\institution{University College London}
}
\email{i.sergey@ucl.ac.uk}

\author{Amrit Kumar}
\affiliation{
\institution{National University of Singapore}
}
\email{amrit@comp.nus.edu.sg}

\author{Aquinas Hobor}
\affiliation{
\institution{Yale-NUS College \\ National University of Singapore}
}
\email{hobor@comp.nus.edu.sg}

\begin{abstract}

This paper outlines key design principles of \scilla---an intermediate-level
	language for verified smart contracts.

\scilla provides a clean separation between the communication aspect of smart
	contracts on a blockchain, allowing for the rich interaction patterns, and
	a programming component, which enjoys principled semantics and is amenable
	to formal verification.
\scilla is not meant to be a high-level programming language, and we are going
	to use it as a translation target for high-level languages, such as
	Solidity, for performing program analysis and verification, before further
	compilation to an executable bytecode.

We describe the automata-based model of \scilla, present its programming
component and show how contract definitions in terms of automata streamline
the process of mechanised verification of their safety and temporal properties.

\end{abstract}

\maketitle

\section{Introduction}

Smart contracts are a mechanism for expressing computations on a blockchain,
\ie, a decentralised Byzantine-fault-tolerant distributed ledger. In addition
to typical state of computations, a blockchain stores a mapping from
\emph{accounts} (public keys or addresses) to quantities of \emph{tokens} owned
by said accounts. Execution of an arbitrary program aka a \emph{smart contract}
is done by \emph{miners}, who run the computations and maintain the distributed
ledger in exchange for a combination of \emph{gas} (transaction fees based on
the execution length, denominated in the intrinsic tokens and paid by the
account calling the smart contract) and \emph{block rewards} (inflationary
issuance of fresh tokens by the underlying protocol). One distinguishing
property of smart contracts, not found in standard computational settings is
the transfer of tokens between accounts.

%

One of the challenges of writing smart contracts is that the
implemented operational semantics of smart contract languages admit
rather subtle behaviour that diverge from the ``intuitive
understanding'' of the language in the minds of contract developers.
Some of the largest attacks on smart contracts, \eg, the
attack on the DAO~\cite{dao} and Parity wallet~\cite{parity}
contracts, have turned on such divergencies.\footnote{By sending money
  to a user-chosen address, the DAO actually called user-chosen code,
  which in turn executed an unexpected callback into the original
  contract, which was in a ``dirty''
  state~\cite{Sergey-Hobor:WTSC17}.} Software development techniques
that have proven very effective in other domains such as app
development (\eg, ``move fast and break things''~\cite{cve}) have not
translated successfully to smart contract development because it is
nearly impossible to patch a contract once deployed due to the
anonymous Byzantine execution environment of a public
blockchain~\cite{dao}. Moreover, software engineering techniques,
such as static and dynamic analysis tools such as
Manticore~\cite{manticore}, Mythril~\cite{mythril},
Oyente~\cite{oyente}, Solgraph~\cite{solgraph} have not yet proven to
be effective in increasing the reliability of smart contracts.

Formal methods, such as verification and model checking, are an attractive
alternative for increasing the reliability of smart
contracts~\cite{Bhargavan16,securify,pirapira}. Formal methods can provide
precise definitions for operational behaviour, and therefore can illuminate and
hopefully reduce subtle behaviour. Generally speaking, formal methods can
produce more rigorous guarantees about program behaviour: mathematical proofs
instead of summaries of accumulated ad-hoc experience. Moreover, formal methods
can provide static guarantees, guaranteeing safety and liveness properties
\textbf{before contracts are irrevocably committed to the blockchain}.
%


In order to apply formal methods efficiently in such a new setting to reason
about smart contracts and enable efficient language-based
verification~\cite{Sheard-al:FOSER10}, one must weigh several factors:

\begin{itemize}

\item \textbf{Expressivity.} There is a trade-off between making a
  language simpler to understand and making it more expressive.
  Bitcoin script~\cite{bitcoin-script} occupies the ``simpler'' end of
  the spectrum: contracts basically specify validity conditions
  (simple expressions) that must hold before coins can be transferred.
  Ethereum~\cite{eth-yellow} occupies the ``expressive'' end of the
  spectrum, with a Turing-complete instruction set. However,
  expressivity is not free. Turing-complete languages are more complex
  to reason about, especially in an automated manner. Moreover,
  infinite computations are neither possible nor desirable on a
  blockchain due to the use of gas to compensate miners (an infinite
  loop will happily consume as much gas as one cares to feed it even
  if no progress is being made). It is as yet unclear if the
  expressivity of a fully Turing-complete instruction set is necessary
  to support a practical smart contract ecosystem.
  %

\item \textbf{State.} Fundamentally, the blockchain is a stateful
  database due to the necessity of maintaining and securely updating
  the mapping between accounts and tokens owned. Moreover, the
  ``event-driven'' style of programming employed by many smart
  contracts (which tend to wait for messages, act on them, and then
  return to waiting for the next message) requires the storage of
  contract state between calls. A contract implementing an
  \emph{Initial Coin Offering} (ICO) campaign, which records each
  contributor and the size of their contributions is a standard
  example of such a stateful event-driven contract. On the other hand,
  purely functional languages are less error-prone, harder to attack,
  easier to parallelise, and easier to reason about, so there are good
  reasons to consider approaches that use mutable state sparingly.

\item \textbf{Communication.} Contracts are often used to allow
  multiple mutually-distrusting parties to interact. This interaction
  can occur in several ways: by one contract calling another, by
  raising an event (to be seen and handled off-chain), or by off-chain
  computations calling back into the blockchain in later blocks.
  Communication is highly desirable, but can introduce both genuine
  and faux-concurrent behaviour, especially in a Byzantine
  environment, enabling attacks due to potentially corrupted state.

\item \textbf{Meaning of execution.} Operational semantics should be
  clear and principled, minimising the chance of informal
  misunderstanding. Moreover, there should be support for
  machine-checked formal reasoning, ideally both
  automatically-generated (for simpler properties) and human-assisted
  (for more complex ones).

	
\end{itemize}

\noindent
In this work, we present \scilla: a novel intermediate-level
programming language for smart contracts. 
By ``intermediate'' we mean that we do not expect most programmers to
write in \scilla directly, any more than most programmers write in
\texttt{x86} assembly directly. Instead, the typical path will be to
compile a higher-level language to \scilla and then further to an
executable bytecode, very much in a tradition of
optimizing~\cite{PeytonJones:BOOK} and verified
compilers~\cite{Leroy:POPL06}.
\noindent
\scilla aims to achieve both \emph{expressivity} and
\emph{tractability}, while enabling rigorous formal reasoning about
contract behavior, by adopting the following fundamental design
principles, based on separation of programming concerns:

\paragraph{Separation between {computation} and {communication}.~}

Contracts in \scilla are structured as \emph{communicating automata}: every
in-contract computation (\eg, changing its balance or computing a value of a
function) is implemented as a standalone, atomic \emph{transition}, \ie,
without involving any other parties.
Whenever such involvement is required (\eg, for transferring control to another
party), a transition would end, with an explicit communication, by means of
sending and receiving messages.
The automata-based structure makes it possible to disentangle the
contract-specific effects (\ie, transitions) from blockchain-wide interactions
(\ie, sending/receiving funds and messages), thus providing a clean reasoning
mechanism about contract composition and invariants.

\paragraph{Separation between effectful and pure computations.~}

Any in-contract computation happening within a transition has to terminate, and
have a predictable effect on the state of the contract and the execution.
In order to achieve this, we draw inspiration from \emph{functional programming}
with effects, drawing a distinction between pure expressions (\eg, expressions
with primitive data types and maps), impure local state manipulations (\ie,
reading/writing into contract fields) and blockchain reflection (\eg, reading
current block number).
By carefully designing semantics of interaction between pure and impure
language aspects, we ensure a number of foundational properties about contract
transitions, such as progress and type preservation, while also making them
amenable to interactive and/or automatic verification with standalone tools.

\paragraph{Separation between {invocation} and {continuation}.~}

Structuring contracts as communicating automata provides a
computational model, known as \emph{continuation-passing style} (CPS),
in which every call to an external function (\ie, another contract)
can be done as the absolutely last instruction.
While this programming style is helpful for avoiding multiple pitfalls stemming
from interaction between separate contracts (\eg, uncontrolled
reentrancy~\cite{dao}), it might be difficult to program with or use as an
intermediate representation.
To regain the expressivity, while retaining the principled structure of an
automata, we add a special kind of transitions---\emph{continuations}---that
are invoked by the execution environment.
Thanks to the mechanism of explicit
continuations~\cite{Reynolds:HOSC98}, we can provide a straightforward
translation from languages like Solidity to \scilla, yet, keeping the
automata structure as a foundational model for analysis and
verification.

\subsection*{Paper outline}
\label{sec:paper-outline}

In the rest of this manuscript, we will describe main components of
\scilla. In Section~\ref{sec:automata} we present its computational
model, based on communicating automata. In
Section~\ref{sec:verification} we demonstrate the support for reasoning
about properties of contract executions, enabled by the automata
model. 
Future design choices \wrt contract verification are discussed in
Section~\ref{sec:discussion}.
%
%
%
We provide a survey of related contract language design proposals in
Section~\ref{sec:related} and conclude in Section~\ref{sec:conclusion}.


\section{Contracts as Communicating Automata}
\label{sec:automata}

In this section, we explain the key concept of \scilla language design
using a characteristic example---a crowdfunding campaign \`{a} la
Kickstarter.\footnote{\url{https://www.kickstarter.com}} In a
crowdfunding campaign, a project owner wishes to raise funds through
donations from the community.  In the specific example modelled here,
we assume that the owner wishes to run the campaign for a certain
pre-determined period of time.  The owner also wishes to raise a
minimum amount of funds without which the project can not be started.
The campaign is deemed successful if the owner can raise the minimum
goal. In case the campaign is unsuccessful, the donations are returned
to the project backers who contributed during the campaign.

\begin{figure*}[t]
\centering
\begin{tabular}{cc}
  \begin{minipage}{0.50\linewidth}
    {

      \begin{lstlisting}[language=Scilla, numbers=left,mathescape=true,basicstyle=\scriptsize\ttfamily]
contract Crowdfunding 
 (owner     : address,
  max_block : uint,
  goal      : uint)

(* Mutable state description *)
{
  backers : address => uint = [];
  funded  : boolean = false;
}

(* Transition 1: Donating money *)
transition Donate 
  (sender : address, value : uint, tag : string)
  (* Simple filter identifying this transition *)
  if tag == "donate" =>
  bs <- & backers;
  blk <- && block_number; 
  let nxt_block = blk + 1 in
  if max_block <= nxt_block
  then send (<to -> sender, amount -> 0,
	      tag -> "main",
	      msg -> "deadline_passed">, MT)
  else
    if not (contains(bs, sender))
    then let bs1 = put(bs, sender, value) in
         backers := bs1;
         send (<to -> sender,
                amount -> 0,
	        tag -> "main",
	        msg -> "ok">, MT)
    else send (<to -> sender,
                amount -> 0,
	        tag -> "main",
	        msg -> "already_donated">, MT)
\end{lstlisting}
}
\end{minipage}
&
\begin{minipage}{0.45\linewidth}
{
\begin{lstlisting}[mathescape=true,language=Scilla,basicstyle=\scriptsize\ttfamily,numbers=left,firstnumber=36]
(* Transition 2: Sending the funds to the owner *)
transition GetFunds
  (sender : address, value : uint, tag : string)
  (* Only the owner can get the money back *)
  if (tag == "getfunds") && (sender == owner) =>
  blk <- && block_number;
  bal <- & balance;
  if max_block < blk
  then if goal <= bal
       then funded := true;   
            send (<to -> owner, amount -> bal,
                   tag -> "main", msg -> "funded">, MT)
       else send (<to -> owner, amount -> 0,
                   tag -> "main", msg -> "failed">, MT)
  else send (<to -> owner, amount -> 0, tag -> "main",
   	      msg -> "too_early_to_claim_funds">, MT)

(* Transition 3: Reclaim funds by a backer *)
transition Claim
  (sender : address, value : uint, tag : string)
  if tag == "claim" =>
  blk <- && block_number;
  if blk <= max_block
  then send (<to -> sender, amount -> 0, tag -> "main",
              msg -> "too_early_to_reclaim">, MT)
  else bs <- & backers;
       bal <- & balance;
       if (not (contains(bs, sender))) || funded ||
          goal <= bal 
       then send (<to -> sender, amount -> 0,
                   tag -> "main",
	           msg -> "cannot_refund">, MT)
       else 
       let v = get(bs, sender) in
       backers := remove(bs, sender);
       send (<to -> sender, amount -> v, tag -> "main", 
              msg -> "here_is_your_money">, MT)
\end{lstlisting}
}
\end{minipage}
\end{tabular}
\caption{\texttt{Crowdfunding} contract in \scilla: state and transitions.}
\label{fig:kick} 
\end{figure*}

An implementation of the contract in \scilla is given in Figure~\ref{fig:kick}.
The design of the \code{Crowdfunding} contract is intentionally simplistic (for
example, it does not allow the backers to change the amount of their donation),
yet it shows the important features of \scilla, which we elaborate upon.

The contract is parameterised with three values that will remain immutable
during its lifetime (lines 2--4): an owner account address \code{owner} of type
\code{address}, a maximal block number \code{max_block} (of type \code{uint}),
indicating a deadline, after which no more donations will be accepted from
backers, and a \code{goal} (also of type \code{uint}) indicating the amount of
funds the owner plans to raise. The \code{goal} is not a hard cap but rather
the minimum amount that the owner wishes to raise.

What follows is the block of mutable \emph{field declarations} (lines~7--10).
The mutable fields of the contract are the mapping \code{backers} (of type
\code{address} \code{=>} \code{uint}), which will be used to keep track of the incoming
donations and is initialised with an empty map literal \code{[]}, and a mutable
boolean flag \code{funded} that indicates whether the owner has already
transferred the funds after the end of the campaign (initialised with
\code{false}). In addition to these fields, any contract in \scilla has an
implicitly declared mutable field \code{balance} (initialised upon the
contract's creation), which keeps the amount of funds held by the contract.
This field can be freely read within the implementation, as we will demonstrate
below, but can only modified by explicitly transferring funds to other
accounts.


\subsection{Transitions and messages}
\label{sec:model}

The logic of the contract is implemented by three \emph{transitions}:
\code{Donate}, \code{GetFunds}, and \code{Claim}. The first one serves for
donating funds to a campaign by external backers; the second allows the owner
to transfer the funds to its account once the campaign is ended and the goal is
reached; the final one makes it possible for the backers to reclaim their funds
in the case the campaign was not successful.

One can think of transitions as methods or functions in Solidity contracts.
What makes them different from functions, though, is the atomicity of the
computation enforced at the language level.
Specifically, each transition manipulates \emph{only} with the state of the
contract itself, without involving any other contracts or parties. All
interaction with the external world, with respect to the contract, happens
either at the very start of a transition, when it is initiated by an external
message, or at the end, when a message (or messages), possibly carrying some
amount of funds, can be emitted and sent to other parties.

Each transition can be invoked by a suitable message, which should provide a
corresponding \emph{tag} as its component to identify which transition is
triggered. It is enforced at the compile time that tags define transitions
unambiguously. All other components of the message, relevant for the transition
to be executed, are declared as the transition's parameters. 
For instance, the transition \code{Donate} expects the incoming message to have
at least the fields \code{sender}, \code{value}, and \code{tag}. 
What follows in each transition's definition is the \emph{filter}---an
optional clause \code{if e =>}, where \code{e} is a boolean-returning
computation that can involve reading from the components of the
incoming message and the contract's state, deciding whether the
corresponding transition can be taken.
For instance, the transitions \code{Donate} and \code{Claim} only check that
the tag of a message matches that of the transition; the filter of
\code{GetFunds} (line~40) additionally checks that the sender of the message is
the contract's owner.

To keep the logic of filters simple, we deliberately disallow complex
expressions in them, as well as write-interaction with the contract's state.
Any further checks relating to the incoming message with the contract's state
can be implemented in a transition's body, as we describe further.

\subsection{Basics of program flow}
\label{sec:programming-language}

Every transition in a contract can be roughly thought of as a function that
maps an incoming message and an initial contract state to a new contract state
and a set of outgoing messages. Ignoring the state-manipulation aspect for a
moment, let us consider the functional component of \scilla. 

As implementations of transitions in Figure~\ref{fig:kick} demonstrate, the
language syntax includes binding of \emph{pure} expressions (such as arithmetic
and boolean operations, as well as manipulation with mappings) via OCaml-style
\code{let-in} construct (\eg, lines~19 and~69).
Basic control flow also includes branching \code{if-then-else} statements,
whose semantics is standard. 
At the moment, we keep an agnostic view \wrt the pure component of the
language, which will be fixed later and can be as expressive as a polymorphic
lambda-calculus~\cite{Reynolds:PS74,Girard:PhD}.
Furthermore, looping constructs are not present in our working example, but we
are planning to support them via well-founded recursive function definitions,
so their termination can be proved statically.

Every transition's last command, in each of the execution branches, is
either sending a set of messages, or simply returning.
Messages are encoded as vectors \code{<...>} of name \code{->} value
entries, including at least the destination address (\code{to}), an
amount of funds transferred (\code{amount}) and a default tag of the
function to be invoked (\code{tag}).
All transitions of the \code{Crowdfunding} end by sending a message
to either the sender of the initial request or the contract's owner.
For example, depending on the state of the contract and the
blockchain, the transition \code{GetFund} might end up in either
sending a message with its balance to the contract's owner, if the
campaign has succeeded and the deadline has passed, or zero funds with
a corresponding text otherwise.

In addition to a message, the trailing \code{send} command of a
transition includes a \emph{continuation} value to indicate possible
further execution in a return-flow. The \code{Crowdfunding} contract
does not require any such executions, so all continuations are
``empty'' (\ie, they do not initiate any further execution after the
\emph{callee} contract returns), which is indicated by the literal
\code{MT}.
We will provide examples of contracts involving non-trivial
return-flow (and, hence, featuring interesting continuations) in
Section~\ref{sec:cont}.

\subsection{State and effects}
\label{sec:effects}

In addition to performing computations with the components of the
incoming messages and parameters of the contract, every transition can
manipulate with the state of a contract itself, \ie, read/write from/to
its mutable fields, as well as read from the blockchain.

The state of the contract, represented by its fields, is mutable: it can be
changed by the contract's transitions. A body of a transition can \emph{read}
from the fields, assigning the result to immutable stack variables using the
specialised syntax \code{x <- & f;}, where \code{f} is a field name and
\code{x} is a fresh name of a local variable (\eg, lines~17 and~42).
In a similar vein, a body of transition can \emph{store} a result of a pure
expression \code{e} into a contract field \code{f} using the syntax \code{f :=
e;} (as in lines~27 and~70).
The dichotomy between pure expressions (coming with corresponding binding form
\code{let-in}) and impure (``effectful'') commands manipulating the field
values, is introduced on purpose to facilitate logic-based verification of
contracts, reasoning about the effect of a transition to the contract's state,
while abstracting away from evaluation of pure expressions, similarly to how it
is done in functional languages, such as Haskell.

In addition to reading/writing contract state, each transition
implementation can use read-only introspection on the current state of
the blockchain using the ``deep read'' operation \code{x <- && g;},
where \code{g} is a name of the corresponding aspect of the underlying
blockchain state.
For example, the \code{Crowdfunding} contract reads the number of a
current block in lines~18 and~41. 
A syntactic emphasis on the contract's interaction with the
blockchain's current state makes it possible to enable reasoning about
contract liveness properties, spanning its long-term behavior, as we
will show in Section~\ref{sec:verification}.

At the moment, the model of \scilla does not feature explicit
exceptions, as those are going to be implemented at the level of
runtime, without a possibility of raising in the code.\footnote{This
  design choice might change in the future, in favor of implementing
  exceptions as another computational effect.}




\subsection{Advanced control flow and continuations}
\label{sec:cont}

It is not uncommon for a contract to call another contract, for instance,
implementing a library, and then use the result of the call in the rest of the
execution. Currently, the main model of computation in \scilla prevents this,
as it corresponds to a \emph{tail-call} program form: passing control to
another contract can be done only by explicitly sending it a message, not via a
call \emph{within} the transition. Existing high-level languages for contracts,
such as Solidity, allow \emph{non-tail} calls from the middle of a contract
execution, and require the notion of program stack in order to handle the
returned result. The presence of non-tail calls in Solidity is what enabled the
infamous DAO exploit~\cite{dao}, which was due to the fact that the rest of a
contract computation, setting the fields accounting for the balance of a
contributor, was performed only upon returning from a call to another contract,
not before. This led to tail-call programming being advocated as a ``good
programming practice'' for smart contracts written in Solidity-like
languages~\cite{consensys}.

While the core language of \scilla enforces non-tail calls from contract
transitions, we acknowledge the need for them in certain applications, and
introduce an additional component to the execution: explicit
\emph{continuations}. Continuations can be thought of as ``the rest of
computation'', to be invoked after execution of a call to an external function,
being passed the result of the latter. One can also encode exception handling
via continuations, which, in that case, would play the role of
\ccode{catch}-clauses.

Functional programming languages, such as Haskell and OCaml allow for
encoding continuations as closures (first-class anonymous functions),
thus implementing a style of programming, known as
Continuation-Passing Style (CPS)~\cite{Appel:CPS92}. In programming
languages without first-class functions, such as Solidity,
continuations still can be encoded via a dedicated data type that
``enumerates'' all possible shapes of ``remaining computations'' and a
helper function that gives them an operational meaning, \ie,
``executes'' the continuation.
A transformation of a program from a non-tail-call form to CPS to a
form with continuations encoded as a data type (known as a
\emph{defunctionalisation}~\cite{Reynolds:HOSC98,Danvy-Nielsen:PPDP01})
is a well-studied topic in the research area of compiler
implementation~\cite{Danvy:ICFP08}, and we adopt it in the design of
\scilla.

Specifically, in Figure~\ref{fig:kick}, every tail call made via \scode{send}
also takes a second argument, continuation \code{MT}. This is a constant,
``empty'', continuation (hence the name), which indicates that no remaining
computation should take place after the execution of a callee contract.
However, instead of empty continuation, we could have specified, for instance,
a continuation that expects a callee contract result to return a result of type
\code{uint}, and sends it back in a message to the owner of a \emph{caller}
contract:

\begin{lstlisting}[language=Scilla,mathescape=true,basicstyle=\small\ttfamily]
(* Specifying a  continuation in a Caller contract *)  
continuation UseResult (res : uint)
  send (<to -> owner, amount -> 0,
         tag -> "main", msg -> res>, MT)

(* Using a continuation in a transition of Caller *)
transition ClientTransition 
  (sender : address, value : uint, tag : string)
  (* code of the transition *)
  send (<to -> sender, amount -> 0,
         tag -> "main", msg -> res>, UseResult)

(* Returning a result in a callee contract  *)
transition ServerTransition 
  (sender : address, value : uint, tag : string)
  (* code of the transition *)
  return value
\end{lstlisting}

That is, in a \scilla contract, a continuation is very similar to a transition,
except it takes not a message, but a value, which is, in its turn, returned by
a callee contract using a \code{return} command, which is an alternative to
\code{send} and does not send a message.
The uniformity of contract-specified continuations and transitions makes it
possible to reason about them in the same way as about transitions, in the
style demonstrated further in Section~\ref{sec:verification}.

What makes continuations different from transitions is that they are
``passive'', \ie, their invocation is handled by the semantics of the
execution environment, which maintains them in a stack, invoking the
topmost continuation after the current contract executes a transition
until the \code{return} statement. In contrast, transitions are
``active'', \ie, they should be explicitly invoked, externally or by
other contracts, by sending a message.

Notice that a continuation can itself ``schedule'' another continuation for
later execution as a rest of computation. For instance, in the code snippet
above the \ccode{UseResult} continuation ends with sending a message and
scheduling an \ccode{MT} continuation.  In principle, a continuation can even
schedule itself, which would mean a potentially non-terminating execution. That
is, the only non-well-founded recursion in \scilla can be implemented via
contracts calling themselves or other contracts in a circular way. Such
potentially non-terminating computations involving multiple transitions and
continuations are going to be handled using the \emph{gas} mechanism, similar
to the one implemented in Ethereum. In contrast, standalone transitions of a
contract \emph{always} terminate.

In the future, we are going to implement an automatic translator from
a subset of Solidity into \scilla by employing CPS transformation and
defunctionalisation, in order to produce \scilla contracts formulated
in terms of transitions for direct control flow and continuations for
return-flow.


\section{Mechanised Verification of \scilla Contracts}
\label{sec:verification}

We now turn our attention to \emph{formally reasoning} about executable
contracts implemented in \scilla. The language's automata-based model allows us
to state a formal semantics of a contract's executions, both independent and
parameterised via interaction scenarios with other contracts, as well as to
rigorously capture the properties of a contract's executions during its
life-cycle.
Below, we show how to reason both about safety (\emph{nothing goes wrong}) and
liveness (\emph{certain things may eventually happen}) properties.

We are developing \scilla hand-in-hand with the formalisation of its
semantics and its embedding into the Coq proof
assistant~\cite{Coq-manual}---a state-of-the art tool for mechanised
proofs about programs, based on advanced dependently-typed theory and
featuring a large set of mathematical libraries.\footnote{The
  mechanised embedding of a subset of \scilla into Coq is publicly
  available for downloads and experiments:
  \url{https://github.com/ilyasergey/scilla-coq}.}
In the past, Coq has been successfully applied for implementing
certified (\ie, fully mechanically verified)
compilers~\cite{Leroy:POPL06}, operational
systems~\cite{Gu-al:OSDI16}, concurrent~\cite{Sergey-al:PLDI15} and
distributed applications~\cite{Lesani-al:POPL16}, including
blockchains~\cite{Pirlea-Sergey:CPP18}.

Further in this section, we will show the translation from \scilla to Coq (which
is mostly straightforward), as well as the definition of contract protocols,
semantics and safety/liveness properties, along with the corresponding proof
machinery. We will then present and discuss a series of properties of the
\code{Crowdfunding} contract from Section~\ref{sec:automata} that we have
verified.
In this manuscript, we outline a preliminary simplified model of
contracts, which does not feature resource semantics (\ie,
Ethereum-style ``gas''), full-fledged blockchain reflection, and
advanced control-flow features, such as continuations and exceptions.
All these aspects are orthogonal to the automata-based model we are
describing; that can and will be modelled in the future in our
framework for formal verification, by enhancing the Coq model of the
\scilla contracts and its semantics.

\subsection{Contracts in Coq: basic definitions and properties}
\label{sec:prot-model-contr}

\begin{figure}[t]
\centering


\begin{lstlisting}[language=Coq, basicstyle=\small\ttfamily]
Structure message := 
  Msg { val : value; 
         sender : address; 
         method : tag; 
         body : payload }.
Structure cstate (S : Type) := 
  CState { my_id : address; 
            balance : value; state : S }.
Structure bstate := 
  BState { block_num : nat; 
            (* More components of the blockchain. *) }.

Structure transition := 
  CTrans { ttag : tag; 
            tfun : address -> value -> S -> message -> 
                   bstate -> (S * option message); }.

Structure Contract (S : Type) := Contr {
      (*Account id *)
      acc : address;
      (* Initial balance *)
      init_bal : nat;
      (* Initial state of the contract protocol *)
      init_state : S;      
      (* Contract comes with a set of transitions *)
      transitions : seq (transition S);
      (* All transitions have unique tags *)
      _ : uniq (map ttag transitions) }.
\end{lstlisting}  

\caption{Basic definitions: states, transitions,  and contracts.}
\label{fig:protocols}
\end{figure}

We present a simplified version of \scilla's semantics in Coq. Coq's
programming component, Gallina, is an ML-family
language~\cite{Milner-al:-ML90} with a similar syntax, featuring
first-class functions, algebraic data types, and records. All data
types are immutable, and so are function parameters and local
variables, bound via \ccode{let} keyword. Gallina is intentionally
non-Turing complete: it does not have an implicit state or pointers;
all loops are provably terminating and can be expressed via
\emph{well-founded} recursive functions.

Before translating our first \scilla contract to Coq/Gallina, we first present
the structure of the contract encoding, defined in terms of the automata
model. Our embedding of \scilla to Coq is, thus,
\emph{shallow}~\cite{Garillot-Werner:TPHOLs07}: we model the semantics of the
pure (\ie, non message-passing) component of \scilla contracts via native
Gallina functions, while the message-passing, state-transition aspect of
\scilla is accounted for by encoding contract semantics as a \emph{relation} in
Coq. In essence, this is similar to implementing a regular domain-specific
language with a tailored execution runtime on top of Lisp or Scala.

The top part of Figure~\ref{fig:protocols} shows Coq definitions of the main
concepts constituting \scilla-like contracts. Three data types represented by
Coq's \ccode{Structure} definitions: \ccode{message}, \ccode{cstate}, and
\ccode{bstate} define structures for modelling contract messages, contract
states and blockchain state.
In our model, messages carry four fields: \ccode{value}, \ie, some amount of
funds (isomorphic for now to a natural number of type \code{nat}), and address
of the message sender analogous to Ethereum's \scode{sender} field, a
\ccode{tag} indicating a transition of a callee contract to be invoked, and a
\code{body} message modelling application-specific payload, containing all
remaining fields of the message.
Generic contract state data type \ccode{cstate}, parameterised by an
application-specific state data type \ccode{S}, additionally contains two
mandatory fields, \ccode{my_id} and \ccode{balance}, storing the contract's
(immutable) address and a current balance, which might change during the
contract's lifetime. All additional fields are encapsulated in the
\ccode{state} component, which is to be defined by the contract implementor.
Finally, the \ccode{bstate} data type is used to model the blockchain component
that a contract can read from. For simplicity, the figure omits all components
but the current latest block number, which is represented by a natural number.
Indeed, in the future we are planning to enrich this state for reasoning about
contracts that reflect on the blockchain state, \eg, read from the results of
previous transactions or even the state of other contracts.

The data type \ccode{transition} describes contract transitions: its
\ccode{ttag} component serves to uniquely identify a transition and implement
the message-based dispatch (using \code{method} field of an incoming message),
while \code{tfun} implements the \emph{transition function}, which takes as
arguments the contract's own \ccode{address}, its current balance of type
\ccode{value}, state, as well as an incoming message and a current blockchain
snapshots, and returns a new state and an \emph{optional} message.
No returned message corresponds to an execution step resulting in an exception.

The contract data type \ccode{Contract} serves to package together contract's
transitions, its address, balance, and initial state, and can be thought of as
a template for contract definitions in Solidity-like languages. Most of the
fields of the \ccode{Contract} record (parameterised with the
application-specific contract state type \ccode{S}) are, thus,
self-explanatory. 
We represent the fixed collection of transitions of a contract automata by a
\ccode{seq}uence of values of type \ccode{transition}, \ie, tagged transfer
functions. 
The only unusual component of a contract definition is its last \emph{property}
field. It intentionally has no name, which is indicated by the placeholder
\ccode{_}. What is important, however, is its \emph{type}, which asserts a
statement. Specifically, it requires that all transitions tags are
\ccode{uniq}ue.
Intuitively, this is a basic \emph{contract validity property}, which our
encoding enforces at the level of the framework, rather than individual
contract. Unlike Solidity and other object-oriented languages, such as Java and
C\#, taking Coq as a host for domain-specific embedding makes it possible to
statically encode and enforce data structure invariants. In other words, it
will be impossible to construct a contract in our embedding, such that it has
two or more transitions with the same tag.
By design, for now this is the \emph{only} property imposed for any \scilla
contract and verified by a compiler during the embedding into Coq. 
Any other correctness guarantee is contract specific, and will have to be
proven for a particular instance of the \ccode{Contract} data type, \ie, for a
particular user-defined contract.

\subsection{Semantics, safety, and consistency properties}
\label{sec:prot-model-contr}

\begin{figure}[t]
\centering


\begin{lstlisting}[language=Coq, basicstyle=\small\ttfamily]
(* A single-step execution: pre/post states and output; 
   contract-specific state S is now assumed to be fixed. *)
Structure step := 
  Step { pre  : cstate S; 
          post : cstate S; 
          out  : option message }.
Definition trace := seq step.
Definition schedule := seq (bstate * message).

(* In the following definition, a contract automata C 
   is implicit and fixed. *)
Definition step_prot (pre : cstate S) (bc : bstate) 
                      (m : message) : step :=
  let CState id bal s := pre in
  let (s', out) := apply_transition C id bal s m bc in
  let bal' := if out is Some m' 
               then (bal + val m) - val m' else bal in
  let post := CState id bal' s' in
  Step pre post out.

(* Map a schedule into a trace *)
Fixpoint execute (pre : cstate S) (sc: schedule) : trace :=
  if sc is (bc, m) :: sc'
  then let stp := step_prot pre bc m in
       stp :: execute (post stp) sc'
  else [::].

Definition state0 := 
  CState (acc C) (init_bal C) (init_state C).
Definition execute0 sc := 
  if sc is _ :: _ then execute state0 sc 
  else [:: Step state0 state0 None].
\end{lstlisting}  
\caption{Contract traces and semantics.}
\label{fig:semantics}
\end{figure}

Having defined the basic terminology of contract embedding into Coq, we can now
define the \emph{meaning} of a contract's behaviour in the form of execution
traces. 
Figure~\ref{fig:semantics} first describes a data type \code{step}, which
captures a triple, corresponding to single step of a contract taking a
particular transition and modifying its state accordingly: a contract
\emph{pre-state} \ccode{pre}, \emph{post-state} \code{pos} and an optional
output \ccode{out}. A \ccode{trace} of a contract is defined as a sequence of
steps. 
The auxiliary data type \code{schedule}, represented by a sequence of
blockchain states and messages, allows us to model \emph{all} possible
interactions between the contract and its environment (\ie, other contracts and
the constantly changing blockchain).

To see how all contract traces can be modelled by considering all possible
schedules, let us first take a look at the semantic function \ccode{step_prot}
that takes a contract state \ccode{pre}, a blockchain state \ccode{bc} and an
incoming message \ccode{m}, resulting in a \ccode{step} instance. 
The way it is defined, it simply finds an appropriate transition in a contract
definition \ccode{C} and applies it using the function
\ccode{apply_transition}, whose definition we have omitted for brevity. In the
case if no transition matching \ccode{m}'s tag is found, the contract's state
and balance are left unchanged; otherwise the new state \ccode{s'} and an
output are obtained and, together with an updated contract balance \ccode{bal'}
are used to construct the final state \ccode{post}.

That is, for a fixed contract \ccode{C}, every component of the schedule (\ie, a
pair blockchain state, message) determines its next step, and, hence, the changes in
its state and balance. Therefore, given an initial state \ccode{pre} and a
schedule \ccode{sc}, we can define a contract execution as a trace, obtained by
consecutively processing all components of the schedule by the
contract---precisely what is defined by a recursive semantic function
\ccode{execute}.
Finally, given an initial contract state, balance, and account, the valid
execution of a schedule \ccode{sc} is defined via the semantic meta-function
\ccode{execute0}, which executes the entire given schedule \ccode{sc} or simply
performs an identity step from the initial state in the case if \ccode{sc} is
empty.

A trace-based semantics of contracts, provided by means of the definitions of
\ccode{execute} and \ccode{execute0} makes it possible to formulate generic
classes of contract correctness conditions, independently of the specifics of
the end-used contracts or properties of interest.

\paragraph{Safety.~}

We first define a predicate \ccode{I} on a contract state (denoted, in Coq
terms, by a ``function type'' \ccode{cstate S -> Prop} from the type of states
\ccode{cstate S} to propositions \ccode{Prop}) to be a \emph{safety property}
if it holds at any state of a contract, that can be obtained as a result of
interaction between the contract and its environment, starting from the initial
state. The following Coq definition states this formally:

%
\begin{lstlisting}[language=Coq, basicstyle=\small\ttfamily]
Definition safe (I : cstate S -> Prop) : Prop :=
  (* For any schedule sc, pre/post states and out... *)
  forall sc pre post out, 
  (* such that the triple Step (pre, post, out) 
      is in the trace obtained via sc *)
  Step pre post out \In execute0 sc -> 
  (* both pre and post satisfy I *)
  I pre /\ I post.
\end{lstlisting}

A safety property means some universally true correctness condition holds at
any contract's state, which is reachable from its initial configuration via
\emph{any} schedule \ccode{sc}. Typical examples of safety properties of
interest include: ``a contract's balance is always positive'', ``a contract's
balance equals the sum of balances of its contributors'', or ``at any moment no
money is blocked on the contract''.
The definition above thus defines safety by universally quantifying over
\emph{all} schedules \ccode{sc}, as well as step-triples \ccode{Step pre post
out} that occur in a trace, obtained by following \ccode{sc}.
While this definition is descriptive, it is not very pleasant to work with.
This is why we define a natural \emph{proof principle} to establish safety of a
property \ccode{I}. The proof principle is stated as the following Coq's
\ccode{Lemma}:\footnote{One can think of lemmas as of ``library functions'',
whose statement (type) is of importance but the proof (implementation) is opaque
and can be ignored by the users.}


\begin{lstlisting}[language=Coq,basicstyle=\small\ttfamily]
Lemma safe_ind (I : cstate S -> Prop) :
  (* (a) *) 
  I state0 ->
  (* (b) *) 
  (forall pre bc m, (method m \in tags p) -> I pre -> 
                  I (post (step_prot pre bc m))) -> 
  (* (conclusion) *)
  safe I. 
\end{lstlisting}

That is, in order to show that the property \ccode{I} is indeed a
safety one, one has to show that (a) it holds in the initial
contract's state \ccode{state0}, \emph{and} (b) if it holds in a
pre-state \ccode{pre} (\ie, \ccode{I pre}), and a contract makes a
transition in a blockchain state \ccode{bc} via a message \ccode{m},
then \ccode{I} holds over the post-state: \ccode{I (post (step_prot
  pre bc m))}.

The \emph{proof} of this lemma, justifying the validity of this
induction principle, is in our Coq sources and is omitted from this
document for the sake of brevity. 
One can notice that the lemma \ccode{safe_ind} is strikingly similar to
the classical notion of mathematical induction for natural numbers, and
it is indeed, an induction on a number, namely, the \emph{length} of a
schedule and, correspondingly a contract execution trace.
This is not the only possible induction principle one can adopt for
proving safety of a contract with respect to a given property.
For instance, certain other properties are easier to be proven by
induction on the size of a certain contract field, considering a trace
arbitrary, but fixed.
Fortunately, almost any valid proof principle for safety can be
encoded in Coq as a lemma, similar to the one above.
Our plans involve developing a library of multiple proof principles
for proving diverse contract properties, formulated as Coq lemmas,
along with the comprehensive documentation on their usage and
applicability, so the contract developer could pick one depending on
her goal and on the nature of the property or a contract.

\paragraph{Temporal properties.~}

Sometimes a property of interest cannot be expressed in terms of a
predicate over a single state, as it describes entire sub-traces of a
contract execution. 
Such properties are traditionally expressed using connectives of
\emph{Temporal Logic}~\cite{Pnueli-FOCS77}, that relate two or more
states in a trace, generated by executing a state-transition system,
such as \scilla contract.

Reasoning principles and the corresponding connectives customary for
temporal logic can be encoded in Coq by means of defining logical
higher-order operators on traces, in the spirit of higher-order
functions in programming languages such as OCaml or Haskell.
We are still in the process of determining a minimal set of such
connectives, necessary to declaratively describe the contract
behaviour. 
As an example, let us consider a temporal connective
\ccode{since_as_long p q r}, which means the following: once the
contract is in a state \ccode{st}, in which ($i$) the property
\ccode{p} is satisfied, each state \ccode{st'} reachable from
\ccode{st} ($ii$) satisfies a binary property \ccode{q st st'} (with
respect to \ccode{st}), as long as ($iii$) every element of the
schedule \ccode{sc}, ``leading'' from \ccode{st} to \ccode{st'}
satisfies a predicate \ccode{r}.

The corresponding Coq encoding of the \ccode{since_as_long} connective
is given below. We first specify reachability between states
\ccode{st} and \ccode{st'} via a schedule \ccode{sc} as the state
\ccode{st'} being the \emph{last} post-state in a trace obtained by
executing the contract from \ccode{st} via \ccode{sc}:
\begin{lstlisting}[language=Coq,basicstyle=\small\ttfamily]
Definition reachable (st st' : cstate S) sc := 
  st' = post (last (Step st st None) (execute st sc)).
\end{lstlisting}
We next employ the definition of reachability to define the
\ccode{since} connective, which is parameterised by predicates
\ccode{p}, \ccode{q} and \ccode{r}. The premises ($i$)--($iii$) are outlined
in the corresponding comments in the following Coq code:
\begin{lstlisting}[language=Coq,basicstyle=\small\ttfamily]
(* q holds since p, as long as schedule bits satisfy r. *)
Definition since_as_long (p : cstate S -> Prop) 
                            (q : cstate S -> cstate S -> Prop)
                            (r : bstate * message -> Prop) :=
  forall sc st st', 
    (* (i) st satisfies p *)
    p st -> 
    (* (ii) st' is reachable from st via sc *) 
    reachable st st' sc -> 
    (* (iii) any element b of sc satisfies r *) 
    (forall b, b \In sc -> r b) -> 
    (* (conclusion) q holds over st and st' *) 
    q st st'. 
\end{lstlisting}
Why this logical connective is useful for reasoning about contract correctness?
As we will show further, it makes it possible to concisely express
``preservation'' properties relating contract balance and state, so that they
hold as long as certain actions do not get triggered by some of the contract's
users.

\subsection{Embedding \scilla into Coq}
\label{sec:transl-from-scilla}

\begin{figure*}[t]
\centering
\begin{tabular}{cc}
\begin{minipage}{0.50\linewidth}
{
\begin{lstlisting}[mathescape=true,language=Coq,basicstyle=\scriptsize\ttfamily,numbers=left]
(* Contract-specific state S *)
Structure crowdState := CS {
   owner : address;
   max_block : nat;
   goal : value;
   backers : seq (address * value);
   funded : bool }.

(* Initial parameters *)
Parameter init_owner : address.
Parameter init_block : nat.
Parameter init_goal : value.

(* Initial state *)
Definition init_state := 
  CS (init_owner, init_block, init_max_goal) [::] false.

(* Transition: tag and a transfer function. *)
Definition donate_tag := 1.

Definition donate_fun := fun id bal s m bc =>
  if method m == donate_tag then
    let bs := backers s in
    let nxt_block := block_num bc + 1 in
    let from := sender m in
    if get_max_block s <= nxt_block
    then (s, Some (Msg 0 id from 0 no_msg))
    else if all [pred e | e.1 != from] bs
         (* new backer *)
         then let bs' := (from, val m) :: bs in
              let s'  := set_backers s bs' in
              (s', Some (Msg 0 id from 0 ok_msg))
         else (s, Some (Msg 0 id from 0 no_msg))
  else (s, None).

Definition donate := CTrans donate_tag donate_fun.
\end{lstlisting}
}
\end{minipage}
&
\begin{minipage}{0.45\linewidth}
{
\begin{lstlisting}[mathescape=true,language=Coq,basicstyle=\scriptsize\ttfamily,numbers=left,firstnumber=37]
Definition getfunds_tag := 2.
Definition getfunds_fun : tft := fun id bal s m bc =>
  let from := sender m in
  if (method m == getfunds_tag) && (from == get_owner s) then
    let blk := block_num bc + 1 in
    if get_max_block s < blk
    then if get_goal s <= bal
         then let s' := set_funded s true in
              (s', Some (Msg bal id from 0 ok_msg))
         else (s, Some (Msg 0 id from 0 no_msg))
    else (s, Some (Msg 0 id from 0 no_msg))
  else (s, None).
Definition get_funds := CTrans getfunds_tag getfunds_fun.

Definition claim_tag := 3.
Definition claim_fun := fun id bal s m bc =>
  let from := sender m in
  if method m == claim_tag then
    let blk := block_num bc in
    if blk <= get_max_block s
    then (* Too early to ask for reimbursements! *)
      (s, Some (Msg 0 id from 0 no_msg))
    else let bs := backers s in
         if (funded s) || (get_goal s <= bal)
         (* Cannot reimburse: campaign succeeded *)
         then (s, Some (Msg 0 id from 0 no_msg))
         else let n := seq.find [pred e | e.1 == from] bs in
              if n < size bs
              then let v := nth 0 (map snd bs) n in
                   let bs' := filter [pred e | e.1 != from] bs in
                   let s'  := set_backers s bs' in
                   (s', Some (Msg v id from 0 ok_msg))
              else (* Didn't back or have already claimed *)
                (s, None)
  else (s, None).
Definition claim := CTrans claim_tag claim_fun.
\end{lstlisting}
}
\end{minipage}
\end{tabular}
\caption{\texttt{Crowdfunding} contract translated into Coq:
  contract-specific state, initial parameters, and transitions.}
\label{fig:kick-coq} 
\end{figure*}

Figure~\ref{fig:kick-coq} shows the translation of the \scilla
implementation of the \scode{Crowdfunding} contract from
Figure~\ref{fig:kick} to Coq. 
The translation is mostly straightforward, and for now has been done
by hand, but in the future we intend to automate it. 
We only outline a few discrepancies between the \scilla code and its
Coq counterpart, which were introduced to streamline the reasoning and
make full use of Coq's programming component Gallina. 

The state of the contract is defined using Coq's \ccode{Structure}
declaration, familiar from the previous sections. Unlike the code in
Figure~\ref{fig:kick}, the definition does not distinguish between
immutable contract parameters and mutable fields. However, while we
provide \emph{getters} for all five components of the state (\eg,
\ccode{get_owner}, \ccode{get_backers}, \etc), we provide
\emph{setters} only for the two last fields, \ie, \ccode{set_backers}
and \ccode{set_funded}, which are considered mutable.
For now, we model mappings by associative sequences, hence the
``field'', \ccode{backers} is encoded as a sequence of pairs
\ccode{(address, value)}, indicating the backers and the corresponding
amount they have donated.

The contract's \emph{owner}, maximal block determining the end of the
campaign, and the funding goal are expressed via Gallina's
\ccode{Parameter}s and can be instantiated later, once a specific
contract instance is created. Having those three parameters abstract,
we define the constructor \ccode{init_state} for the initial state
that also instantiates backers with an empty sequence \ccode{[::]},
and sets the boolean \ccode{funded} flag to \ccode{false}.

The remaining Coq code defines the encoding of the three transitions
of the contract, by means of specifying their tags (\eg,
\ccode{donate_tag}), transfer functions (\eg, \ccode{donate_fun}), and
packaging them together in a single transition (\eg, \ccode{donate}).
In this version of the encoding, we model implicit mutable state of the
contract by explicit functional \emph{state-passing style}.
That is, each transition's transfer function takes the current
contract's state \ccode{s}, as well as the incoming message \ccode{m}
and a blockchain state \ccode{bc} as its parameters.
The mapping between \scilla primitive commands is then as follows.
Reading from a contract's state \scode{x <- & f;} is translated into
\ccode{let x := get_f s in ...}. Writing into a contract field
\scode{f := e} is translated into \ccode{let s' := set_f s e in ...}, where
\ccode{s'} is the new, modified state to passed further in the
computation. Finally, reading from the blockchain \scode{x <- && g;}
is encoded via \ccode{let x := get_g bc}. 
A specific shape of a blockchain getter depends on the property being
read. For instance, the current block number can be obtained via
\ccode{block_num bc}, as shown on line 24 of
Figure~\ref{fig:kick-coq}.
Following the shallow embedding style, in our translation, we have
also formulated \scilla's transition filters via Gallina's native
\code{if-then-else} construct.

Other difference between \scilla representation and Coq translation
involves representing mappings: instead of using language-provided
primitives \scode{put/get/contains}, we implement them using Coq's
functions for sequence manipulation. For instance, the call to
\scode{contains(bs, sender)} from Figure~\ref{fig:kick}'s line 25 is
implemented in Gallina translation via \ccode{all [pred e | e.1 != from] bs} in
line 28 of Figure~\ref{fig:kick-coq}, checking that no single entry \ccode{e}
has \code{from} as its first component, that is, making sure that the sequence
does not yet contain a record of \ccode{sender m}.
In a similar spirit, \scode{put} is encoded via appending a head to a sequence
(line 30 of Figure~\ref{fig:kick-coq}), and removing an entry is done via
\ccode{filter} function (line 66 of Figure~\ref{fig:kick-coq}).
Relying on Coq's support for sequence for encoding mappings makes it possible
to reuse a rich library of lemmas about them, thus, sparing us the expense of
having to implement a new library for mappings.\footnote{Although we might
consider implementing such a library in the future to streamline the
translation.}

For simplicity, at the moment our encoding does not account for
explicit reentrancy, \ie, it does not involve continuations, and every
execution branch of every transitions terminates by providing a new
state (which can be equal to the previous state), as well as an
optional output message, to be sent to its destination (third argument
of the \ccode{Msg} constructor). The transitions resulting in
\ccode{None} as an output should be interpreted as resulting in an
exception, which is customary in the functional programming
tradition~\cite{Wadler:FPCA85}. In this case, the state of the contract is left
unchanged.

\subsection{Reasoning about contract behaviour}
\label{sec:reas-about-contr}

A definition of a contract as a state-transition system and modelling
all its execution traces makes it possible to verify its complex
properties \emph{in isolation}, \ie, modularly and independently of
other contracts and users that might interact with it. This is
achieved by defining safety and temporal properties as universally
quantified over schedules, which serve as an external
\emph{oracle}~\cite{Hobor-al:ESOP08} and thus account for all
non-determinism of distributed on-chain interaction. Proving a
property \emph{for all} schedules means proving it for any potential
adversary, as all its interaction with the contract are limited by
what the contract allows for, in terms of its transitions.

We now show how the combination of notions of safety and temporal
properties presented above allows us to verify a contract, proving
that all its behaviours satisfy a certain complex interaction
scenario.
Specifically, for our \ccode{Crowdfunding} example, let us prove that,
once a donation \ccode{d} has been made by a backer with an account
address \ccode{b}, given that the campaign eventually fails, the
backer \ccode{b} will be always able to get their donation \ccode{d}
back. 
There are indeed multiple ways to express this property formally in
terms of contract behaviours. For simplicity, we break the statement
of interest into three independent components, which, in conjunction
provide the requirement stated above.

\paragraph{Property 1: The contract does not leak funds unless the
  campaign has been funded.~~}

First, let us state and prove that the contract does not spend the
money given to it by the backers, unless the campaign has been
funded. To do so, we will make use of Coq's native functions
\ccode{map}, used to apply a function to all elements of a sequence,
and \ccode{sumn} for summing up contents of a sequence.
The property of interest, dubbed \ccode{balance_backed}, is as follows:
\begin{lstlisting}[language=Coq,basicstyle=\small\ttfamily]
Definition balance_backed st : Prop :=
  (* If the campaign has not been funded... *)
  negb funded (state st) ->
  (* the contract has enough funds to reimburse all. *)
  sumn (map snd (backers (state st))) <= balance st.
\end{lstlisting}
For an arbitrary contract state \ccode{st}, it asserts that if the
\ccode{funded} flag is still \ccode{false} in \ccode{st} (\ie,
\ccode{negb funded (state st)}), then the balance of the contract
(\ccode{balance st}) is at least as large as the sum of all donations
made by the recorded backers (\ccode{sumn (map snd (backers (state
  st)))}).

Does this property hold for every state of an arbitrary instance of
the \ccode{Crowdfunding} contract?
To show this, we state the following theorem,\footnote{\ccode{Theorem}
  declarations in Coq are no different from \ccode{Lemma}s, but are
  usually considered more important.} claiming that
\ccode{balance_backed} is a safety property:
\begin{lstlisting}[language=Coq,basicstyle=\small\ttfamily]
Theorem sufficient_funds_safe : safe balance_backed.
\end{lstlisting}
The statement of the theorem relies on the definition of \ccode{safe},
instantiating it (implicitly) for an instance of the
\ccode{Crowdfunding} contract, as well as (explicitly) for the
property \ccode{balance_backed}.
A machine-checked proof of the theorem, which takes less than 50 lines
of Coq code, is conducted by using the induction principle
\ccode{safe_ind} defined above. For instance, the property
\ccode{balance_backed} clearly holds in the initial state of the
contract, as the set of backers is empty (\ccode{[::]}), and the
balance cannot be negative. Thus, we can formally prove that a
non-funded campaign does not lose/spend money of its backers.

\paragraph{Property 2: The contract preserves records of individual donations.~}

So far, we have proved that the non-funded Crowdfunding contract does
not lose money, but what about individual backers? What if the contract
takes the donations and silently transfers them from one backer to
another, or, even worse, removes the backer from its records,
``pretending'', that no donation has ever been made. 
To assert that this is not the case, we rely on the temporal
connective \ccode{since_as_long} defined above and state that, once a
backer made a donation, the record of it is not going to be lost by
the contract, \emph{as long as} the backer makes no attempt to
withdraw its donation.

We first define two auxiliary predicates, specific to our contract and
the shape of its state:
\begin{lstlisting}[language=Coq,basicstyle=\small\ttfamily]
(* Contribution d of a backer b is recorded 
   in the field 'backers'. *)
Definition donated b (d : value) st := 
  get (backers (state st)), b) == [:: (b, d)].

(* b doesn't claim its funding back *)
Definition no_claims_from b (q : bstate * message) := 
  sender q.2 != b.
\end{lstlisting}
The predicate \ccode{donated} specifies that the backer-recording
field of the contract has the corresponding entry \ccode{(b, d)}.
The predicate \code{no_claims_from} is defined on schedule bits and
ensures that a given schedule component \ccode{q} does not contain a
message from the address \ccode{b}.
Together, we use these two predicates to state the desired temporal
property of the contract:
\begin{lstlisting}[language=Coq,basicstyle=\small\ttfamily]
Theorem donation_preserved (b : address) (d : value):
  since_as_long c (donated b d) 
                    (fun _ s' => donated b d s')
                    (no_claims_from b).
\end{lstlisting}
The mechanised proof of this fact is by induction on the trace
connecting the initial state \ccode{s}, in which a donation record is
observed, and any fixed subsequent state \ccode{s'} which is reachable
from \ccode{s} via schedules, satisfying the predicate
\ccode{no_claims_from}.
The existence of such a proof indeed validates the claim that the
contract does not mess up with records during its execution, unless
the backer tries to claim their donations back.

\paragraph{Property 3: If the campaign fails, the backers can get their
  refund.~}

By now we know that the contract does not lose the donated funds and
keeps the backer records intact. Now we need the last piece: the proof
that if a contract is not funded, and the campaign has failed
(deadline has passed and the goal has not been reached), then any
backer with the corresponding record can get the donation back.

\begin{figure}[t]
  \centering
\begin{lstlisting}[language=Coq,basicstyle=\small\ttfamily]
Theorem can_claim_back id b d st bc:
  (* (a) The backer b has donated d, so the contract holds 
          that record in its state *)
  donated b d st ->
  (* (b) The campaign has not been funded. *)
  negb funded (state st) ->
  (* (c) Balance is small: not reached the goal. *)
  balance st < (get_goal (state st)) ->
  (* (d) Block number exceeds the deadline. *)
  get_max_block (state st) < block_num bc ->
  (* (conclusion) Backer b can get their donation back. *)
  exists (m : message),
    sender m == b /\ 
    out (step_prot c st bc m) = Some (Msg d id b 0 ok_msg).
\end{lstlisting}
\caption{A backer can claim back her funds if the campaign fails.}
  \label{fig:claim}
\end{figure}

We state the property of interest as theorem
\ccode{can_claim_back} in Figure~\ref{fig:claim}.
As its premises (a)--(d), the theorem lists all the assumptions about
the state of the contract that are necessary for getting the
reimbursement. The conclusion is somewhat peculiar: it expresses the
\emph{possibility} to claim back the funds by postulating the
\emph{existence} of a message \ccode{m}, such that it can be sent by a
backer \ccode{b}, and the response will be a message with precisely
\ccode{d} funds in it, sent back to \ccode{b}.
The theorem, whose proof is only 10 lines of Coq, formulates the
property as one single-step, yet its statement can be easily shown to
be a safety property, as it is, indeed, preserved by the transitions,
and, after the funds are successfully claimed for the first time, the
premise (a) of the statement is going to be false, hence the property
will trivially hold.

\paragraph{Putting it all together.~}
\label{sec:putting-it-all}

Together, Properties 1--3 deliver the desired correctness condition of
a contract: \emph{once donated money can be claimed back in the case
  of a failed campaign}.
It is indeed not the only notion of correctness that intuitively
should hold over this particular contract, and by proving it we did
not ensure that the contract is ``bug-free''.
For instance, in our study we focused on backers only, while another
legit concern would be to formally verify that the contract's owner
will be able transfer the cumulative donation to their account in the
case if the campaign is \emph{successful}.

In a similar vein, another desired property would be to ensure that no funds
will be locked on the contract forever as recently happened with the Parity
multi-signature wallet~\cite{parity}: in a finite amount of time either all
backers or the owner will be able to relieve the contract of its funds.
The latter correctness condition is of particular interest, as it falls into
the class of \emph{liveness} properties, stating that ``something good
eventually happen''. 
While we do not show its proof here, our formalisation of contract executions
makes it possible to prove liveness properties.

\section{Discussion}
\label{sec:discussion}

Representing smart contracts as communicating automata enables
multiple opportunities for logic-based verification. It also opens new
challenges with respect to the design of better high-level contract
languages and their expressivity.

\subsection{Towards logic-based reasoning about contracts}
\label{sec:towards-logic-based}

As we have demonstrated, together, the combination of universal safety
properties and multiple temporal properties (including
\emph{liveness}, which we briefly sketch later in this section)
constitute various contract correctness criteria.
Indeed, \emph{the notion of correctness is in
  the eyes of the beholder}: each contract comes with a unique set of
correctness conditions, and there is no single property that captures all
possible notions of ``a contract not going wrong''.
Even more so, a behaviour considered erroneous in one contract (\eg,
reentrancy in Ethereum-style contracts) might be a feature exercised
consciously and without harm in another implementation.
Defining precisely the desired contract properties is an art of
\emph{formal specification} and is outside the scope of this
technical presentation. We believe that having a clean contract
semantics and an expressive logical abstractions is of crucial
importance for formally describing the behaviour of blockchain-based
applications.

Therefore, we consider our mission to provide semantic foundations, as
well as a logical vocabulary for making it possible to formulate and
prove \emph{all} reasonable contract properties, giving \scilla
programmers a toolset to implement, specify, and mechanically verify
the contracts with respect to correctness conditions of interest.

We also believe that prototyping a contract language by encoding its
programs and their semantics in a state-of-the art language with
dependent types and a support for formal machine-assisted proofs, such
as Coq, Agda~\cite{Bove-al:TPHOL09}, F$^\star$~\cite{Swamy-al:ICFP11},
or Idris~\cite{Brady:JFP13} provides a principled way to rigorously
specify and verify the implementation for a large class of notions of
correctness in terms of the programs themselves, rather than their
models~\cite{Sergey-al:POPL18}.

\subsection{On Turing-completeness and contract language design}
\label{sec:turing-compl-contr}

Our host framework Coq's programming language Gallina, which was used
to implement a verifiable version of \scilla by means of shallow
embedding, is \emph{not} Turing-complete: each well-typed Coq
function, when applied to concrete arguments, will terminate in a
finite number of steps.
By defining \scilla and its embedding to Coq in a way that its pure
and state-manipulating in-transition computations are modelled by
Coq's pure total functions, we ensured that \emph{all} \scilla
transitions terminate.
This made it possible to conduct verification by means of symbolic
execution of Coq's expressions with explicit state.
Keeping the programming component of transfer functions strictly
terminating, in the future, we should be able to provide better
support for automating the proofs of safety/temporal properties by
employing third-party tools, such as TLA+~\cite{Lamport:TLA} and
Ivy~\cite{Padon-al:PLDI16}.
In our proof-of-concept contract verification effort, we have not
explicitly accounted for analysis of resources or computational
effects, such as exceptions.
Such effects can be modelled in a similar way, \ie, by explicit
encoding of an effect-passing and cost-counting discipline, or by
engineering a version of an indexed monad to keep track of the effect
in the type system~\cite{Nanevski-al:ICFP08,Ahman-al:POPL17}.

While transitions of \scilla contracts are strictly terminating by
construction, they still might take arbitrarily long to compute, \eg,
in the presence of a large data value passed by a message. Even more,
one is still able to implement potentially non-terminating (in the
absence of limited stack length or gas bounds) computations by making
a contract calling itself, directly or indirectly, via a
message-passing mechanism or explicit continuations. This is similar
to implementing recursion via so-called ``Landin's knot'', \ie, not by
means of the language function calling mechanisms, but through the
(blockchain) state, storing a computation (\ie, a contract).
To be able to statically detect such scenarios before a transaction is
fired, we are going to implement an analysis to deliver precise
parameterised cost boundaries on transition executions.

\section{Related Proposals}
\label{sec:related}

Every blockchain either has a (limited) scripting language for transaction
validation as in say Bitcoin or a general purpose smart contract language as in
Ethereum. In the recent years, several new languages improving upon languages
in Bitcoin and Ethereum  have been proposed by the community.  Each language is
usually designed in the context of a specific underlying blockchain.  Below, we
compare \scilla and some of the existing smart contract languages and the
improvement proposals.

\begin{itemize}

\item \textbf{Typecoin (Bitcoin)}~\cite{typecoin}: Typecoin is a
  logical commitment mechanism built on top of Bitcoin to carry
  logical propositions. The underlying idea in Typecoin is to have a
  transaction carry logical propositions instead of coins. In fact,
  each Bitcoin transaction can be translated into a Typecoin
  transaction where the inputs and outputs become propositions and the
  logic would allow to split or merge inputs. This allows transactions
  to be type-checked before they get committed to the blockchain.
  Since, UTXOs can only be merged or split, the underlying logic is
  linear in nature and is not rich to handle complex states as in
  \scilla.

\item \textbf{Simplicity (Bitcoin)}~\cite{simplicity}: Simplicity is
  the most recent language proposed in the context of Bitcoin. Since
  Bitcoin uses a UTXO model and the state of the system is not as
  complex as in say Ethereum, the language design does not need to
  handle read and write to a global state. As a result, the language
  follows Bitcoin's design of self-contained transactions whereby
  contracts do not have access to any information outside the
  transaction. This further implies that Simplicity does not support
  communicating contracts. \scilla on the other hand manages read and
  write to a shared memory space and is designed for an
  account-based model, where contracts can communicate with each
  other.

\item \textbf{Solidity (Ethereum)}~\cite{solidity}: Solidity is the most
	popular smart contract language today. It is a Turing complete language and
		resembles JavaScript. However, the expressivity of the language has
		introduced avenues for several vulnerabilities in the past.  For
		instance, a class of re-entrancy attacks was shown to be possible due
		to arbitrary interleaving of local state manipulations and external
		calls~\cite{dao}. \scilla restricts the computation model to
		communicating automata and mandates external calls to occur at the end
		of the transition. Also, CPS style explicit passing of return values to
		the caller makes reasoning about programs much easier.  With \scilla,
		we further show how to prove critical safety and liveness properties on
		the contract. As for Solidity, the Turing completeness nature of the
		language makes contracts less amenable to formal verification.

\item \textbf{Bamboo (Ethereum)}~\cite{bamboo}: Among all language proposals,
	Bamboo is the closest to \scilla. Bamboo relies on polymorphic contracts
		where one contract changes to another whenever a state change occurs.
		While, in \scilla the transition function changes the state while the
		transition itself does not change. In fact, Bamboo's morphing of
		contracts can be easily encoded via a state component in \scilla. The
		other difference being that Bamboo does not focus on the verification
		aspect of the contracts.

\item \textbf{Babbage (Ethereum)}~\cite{babbage}: Babbage is a conceptual-level
	design of a smart contract proposed in the context of Ethereum. The design
		adopts a mechanical model of writing contracts as opposed to a textual
		program. But, due to its underlying simplicity and lack of formal
		design semantics, it is hard to compare it with \scilla and assess its
		amenability to formal reasoning.

\item \textsf{F$^\star$} \textbf{embedding (Ethereum):}
	Bhargavan~\etal~\cite{Bhargavan16} provide a framework to analyze and
		verify both the runtime safety and the functional correctness of
		Ethereum contracts written in Solidity by translation to
		\textsf{F$^\star$}, a function programming language aimed at program
		verification. The work however does not present a new programming model
		as in \scilla.

\item \textbf{Viper (Ethereum)}~\cite{viper}: Viper is an experimental
  language proposed in the context of Ethereum to ease the
  auditability of smart contracts. The language does not have
  recursive calling and infinite loops. As a result, one can eliminate
  the need to set an upper bound on gas limits that is known to be
  vulnerable to gas limit attacks. Viper also plans to remove the
  possibility of making changes to the state after any non-static
  calls. The idea being that this will prevent reentrancy attacks.
  \scilla takes a slightly stricter approach where any external call
  has to be the last instruction of a transition function.

\item \textbf{Rholang (RChain)}~\cite{rholang}: While \scilla is based on
	communicating automata, Rholang is based on asynchronous polyadic
	$\pi$-calculus that is best suited to work in a concurrent setting.
	Rholang admits unbounded recursion, while \scilla will only allow bounded
	recursion. The other difference comes from the fact that \scilla mandates
	all external calls to be tail calls. As a result, there is no complex
	interleaving of external calls and local instructions. Hence, analyzing and
	proving safety properties in \scilla becomes much easier than in Rholang.

\item \textbf{Michelson (Tezos)}~\cite{michelson}: Michelson is a purely
	functional stack-based language that has no side-effects. On the other
	hand, \scilla is not purely functional as transitions do affect the
	external state.

\item \textbf{Liquidity (Tezos)}~\cite{liquidity}: Liquidity is a high-level
	language that complies with the security restrictions of Michelson. Because
	of its compatibility with Michelson, a similar comparison with \scilla
	holds.

\item \textbf{Plutus (IOHK)}~\cite{plutus}: Plutus is a language based
  on typed $\lambda$-calculus designed to run transaction validation
  for a blockchain like Bitcoin. As a result, the language is simpler
  that \scilla, which allows contracts to manipulate global state and
  make external calls to other contracts. Also, because of the
  automata based design, \scilla contracts are easily composable.

\item \textbf{OWL (BOSCoin)}~\cite{owl}: OWL for Web Ontology Language
  is being developed in the context of BOSCoin. The underlying computation
  model in OWL is Timed Automata~\cite{Andrychowicz14}, and hence bears some
  similarity with \scilla. However, OWL advocates for pure functions without
  side effects, while \scilla is not purely functional and hence allows complex
  yet cleanly separated interactions with other contracts.
     
\item \textsf{F$^\star$} \textbf{dialect (Zen)}~\cite{zen}: Zen uses a dialect
	of \textsf{F$^\star$} for its smart contract language. The smart contracts
	in Zen are stateless and are functionally pure. This means that there are
	no side effects and no interaction with other contracts. While, this
	removes race conditions and any barriers to parallel execution it also
	severely limits the kind of smart contracts that one can build. For
	instance, multiple transactions involving the same smart contract may not
	be easily parallelised, and may have to be executed in series. In \scilla,
	we understand the need of impure functions and we also understand the
	complexities arising from it. As a result, \scilla's computation model
	cleanly separates local computations and external calls that require
	communication. With a clean separation, it becomes possible to eliminate
	complex and undesirable interleaving of local computations and external
	calls.

\end{itemize}

\section{Conclusion \& Future Work}
\label{sec:conclusion}

In this work, we outlined the design of \scilla---an intermediate
level language for smart contracts. \scilla provides a clear
separation between the communication aspect of a smart contract and
its programming component. The underlying computation model in \scilla
is based on communicating automata. We also presented an embedding of
a \scilla contract to Coq and proved safety and liveness properties of
the contract.
Future work consists in defining a formal grammar and semantics of the language
and implementing \scilla, as well as a developing and verifying a number of
contracts in it, on a real-world blockchain platform.




\paragraph{Acknowledgments.}

This research was partially supported by a grant from the National
Research Foundation, Prime Minister's Office, Singapore under its
National Cybersecurity R\&D Program (TSUNAMi project, No.
NRF2014NCR-NCR001-21) and administered by the National Cybersecurity
R\&D Directorate. Hobor's research was partially supported by a grant
from Yale-NUS College R-607-265-322-121.
Sergey's research was partially supported by EPSRC First Grant
EP/P009271/1.

\setlength{\bibsep}{1.5pt}
\bibliography{ref}

\end{document}